# Spin-valley Hall transport induced by spontaneous symmetry breaking in half-filled zero Landau level of bilayer graphene

M. Tanaka[1], Y. Shimazaki[2], I. V. Borzenets[3], K. Watanabe[4], T. Taniguchi[4], S. Tarucha[5], and M. Yamamoto[5]

[1]Department of Applied Physics, University of Tokyo, Bunkyo-ku, Japan. [2]Institute of Quantum Electronics, ETH Zurich, Zurich, Switzerland. [3]Department of Physics, City University of Hong Kong, Hong Kong. [4]National Institute for Materials Science, Tsukuba-shi, Japan. [5]Center for Emergent Matter Science, RIKEN, Wako-shi, Japan

**Intrinsic Hall conductivity, emerging when chiral symmetry is broken, is at the heart of future low energy consumption devices because it can generate non-dissipative charge neutral current. A symmetry breaking state is also induced by electronic correlation even for the centro-symmetric crystalline materials. However, generation of non-dissipative charge neutral current by intrinsic Hall conductivity induced by such spontaneous symmetry breaking is experimentally elusive.**
**Here we report intrinsic Hall conductivity and generation of a non-dissipative charge neutral current in a spontaneous antiferromagnetic state of zero Landau level of bilayer graphene, where spin and valley contrasting Hall conductivity has been theoretically predicted [1,2]. We performed nonlocal transport experiment and found cubic scaling relationship between the local and nonlocal resistance, as a striking evidence of the intrinsic Hall effect. Observation of such spontaneous Hall transport is a milestone toward understanding the electronic correlation effect on the non-dissipative transport. Our result also paves a way toward electrical generation of a spin current in non-magnetic graphene via coupling of spin and valley in this symmetry breaking state combined with the valley Hall effect.**

Theory has predicted gap opening of clean graphene at the charge neutrality point (CNP) induced by electron-electron interaction [1,3,4]. Such gap opening was also experimentally confirmed in bilayer and trilayer graphene [5-7]. The gap is even enhanced in presence of a perpendicular magnetic field due to enhancement of effective Coulomb interaction, where the CNP correspond to half-filled zero-th Landau level ($\nu = 0$ state). This gapped state is distinct from conventional quantum Hall states observed in semiconductor two-dimensional electron gas. It shows unique properties such as the absence of edge channels [8-15], phase transition driven by a tilted magnetic field or a perpendicular electric field [12,17], and the recently observed long-range spin transport [18,19]. While there has been a theoretical debate about the nature of $\nu = 0$ state

[2,20-25], consensus from these experiments is the layer antiferromagnetic (LAF) state, where spins tend to align ferromagnetically due to the exchange interaction within each layer and antiferromagetically between the layers (Fig.1a)[2,20,21]. Neel vector is in-plane and spins are slightly canted out of plane due to the Zeeman energy.

It is known that breaking of crystal chiral symmetry opens a band gap in graphene and that gapped graphene has valley contrasting intrinsic Hall conductivity, which generates transverse valley current to the injected electronic current. This phenomenon is called the valley Hall effect and has also been observed in graphene [26-28]. In the LAF state, intrinsic Hall conductivity also arises with the gap opening, and similar Hall effect is predicted. Here, chiral symmetry is broken due to the spontaneous symmetry breaking driven by the electron-electron interaction. Because opposite spins occupy opposite layers, sign of effective mass term depends on the spin [1,2]. Instead of the conventional valley current, spin-valley current, which is defined as the difference of valley current between the spins, is expected to flow (fig.1b). This theoretically predicted spin-valley Hall effect [1] has been experimentally elusive. Note that the spin-valley Hall effect allows for coupling of spin and valley degrees of freedom. If we inject opposite current for each spin, in other words inject a spin current, a valley current is expected to flow in its transverse direction (Fig.1c). While spin-orbit interaction is weak in graphene, this spin current / valley current conversion may open a new possibility of electric generation of a spin current in graphene with high efficiency, being combined with the conventional valley Hall effect.

In this study, we observed nonlocal transport in the LAF state, which is assigned to the spin-valley Hall effect as an only possible origin. There were a few previous reports on nonlocal transport in graphene under magnetic field [16,29,30], however, no discussion on the spontaneous symmetry breaking state was made. Using bilayer graphene with dual-gated structure, we found gate dependence of nonlocal transport specific for LAF state and cubic scaling relationship between the local and nonlocal resistance similarly to the previous experiments on the valley Hall effect [26,27], consistent with the scenario of the spin-valley Hall effect.

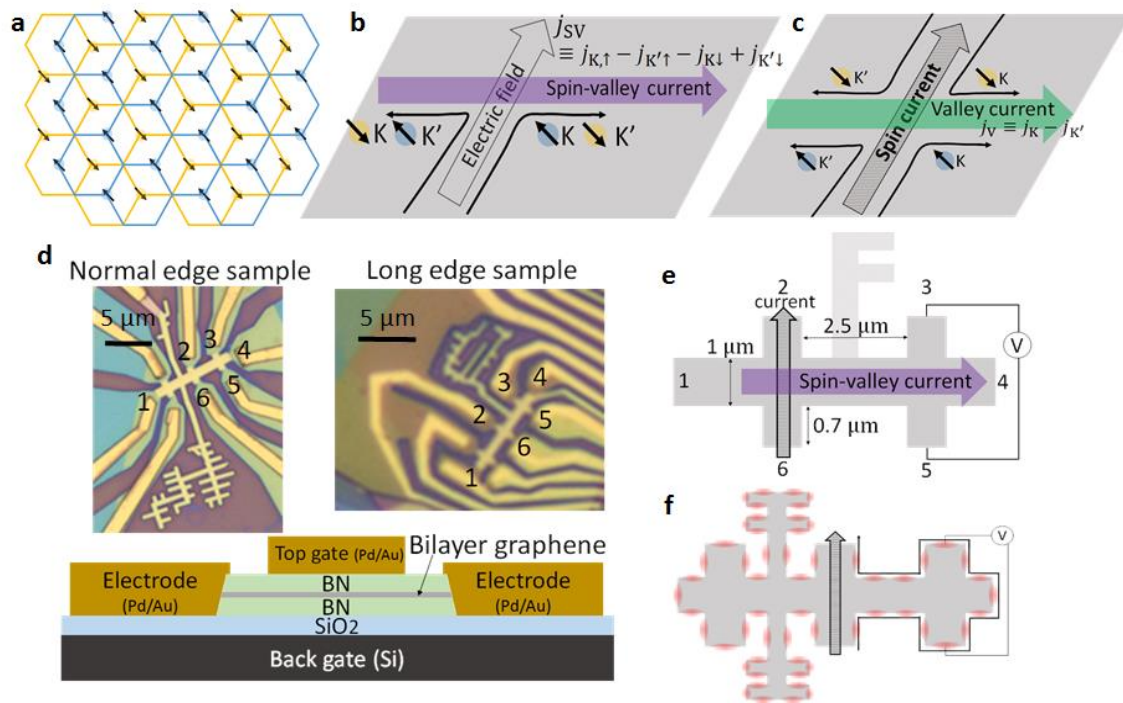

**Fig.1 | Spin-valley Hall effect in LAF state and sample structure.**

**a,** Spin and sublattice configuration of the LAF state at the half-filled $\nu = 0$ Landau level in bilayer graphene. Blue and yellow hexagons are the lattices of top and bottom layers, respectively. Spins shown as arrows tend to align in each layer but become opposite between the layers. **b,** Schematic description of the spin-valley Hall effect and definition of the spin-valley current. Blue and yellow circles indicate electrons of the top and bottom layers, respectively, and arrows indicate spin. In this figure, the valley current flows to the right in the upper layer and to the left in the bottom layer. Net spin current and valley current are both zero in total, but the spin-valley current can become nonzero. **c,** Schematic description of the spin current / valley current conversion. Spin current (=opposite charge current between the layers) generates transverse valley current. **d,** The devices used for our experiments. Upper figure shows the optical microscope image. Two samples have same dimension which is written in Fig. 1e. One side of the long edge sample has a jagged region sticking out of the Hall bar, which is used to examine the contribution of edge transport to the nonlocal transport. The lower figure shows the schematic cross section. Bilayer graphene is encapsulated by high quality hexagonal boron nitride crystals and sandwiched by the top gate and back gate. The thickness of the hexagonal boron nitride is 34 nm (top) and 46 nm (bottom). Electrical contacts to bilayer graphene are taken at the edges. **e,** Schematic description of the nonlocal transport measurement. A charge current is injected between the terminals 2 and 6, and the nonlocal transport mediated by the spin-valley current is measured as the voltage induced between the terminals 3 and 5. **f,** Schematic description of the edge transport as an origin of the nonlocal transport. The voltage drop only occurs along the edge (solid line).

Our experiment is carried out using bilayer graphene (BLG) encapsulated in hexagonal boron nitride (h-BN) (fig.1d). By using the back gate and top gate, we tune the carrier density $n$ and perpendicular electric field (displacement field $D$) independently. The h-BN/BLG/h-BN stack was etched into a Hall bar with the length $L = 2.5$ μm and the width $W = 1$ μm (fig. 1e).

We performed nonlocal transport measurement to demonstrate the spin-valley Hall effect (Fig.1e). It is the scheme widely used to detect the spin Hall effect or the valley Hall effect [26-28]. When the charge current is injected between terminals 2 and 6, the spin-valley current is generated. Because of the non-dissipative nature of spin-valley current in graphene of high crystal quality, it can flow in the longitudinal direction of the Hall bar over a few microns limited mainly by the edge scattering. Voltage is then induced between terminals 3 and 5, owing to the inverse spin-valley Hall effect. Nonlocal resistance ($R_{NL}$) is defined as the ratio of the injected current and detected nonlocal voltage ($V_{3-5}/I_{2-6}$).

We also measured the local resistance ($R_L$) (4-terminal resistance $V_{6-5}/I_{1-4}$) and bias dependence of the two-terminal differential conductance to evaluate the gap size. All data except for that of the temperature dependence was taken at 1.7 K.

Here we shows the top gate and back gate dependence of $R_L$ in Figure 2a-c. At zero magnetic field, $R_L$ at the CNP monotonically increases with the displacement field due to the opening of a single particle band gap. We identified the $D = 0$, $n = 0$ point as the minimum along the CNP line (red circle). When the magnetic field of 2 T is applied, a local maximum appears around the $D = 0$, $n = 0$ point similarly to the previous report [12], indicating the development of LAF insulating state (Fig. 2d). This state becomes more robust for the higher magnetic field.

We performed nonlocal transport measurement for this state. Figure 2e and 2f shows gate voltage dependence of $R_{NL}$ under the magnetic field of 2 T and 4 T. $R_{NL}$ increases around $D = 0$, $n = 0$, where $R_L$ also increases. The value of $R_{NL}$ is much larger than the expected value for the trivial classical current diffusion.

Figure 2g and 2h shows the carrier density dependence of $R_\text{L}$ and $R_\text{NL}$ at $D = 0$ under the magnetic field of 1 ~ 8 T. The peak of $R_\text{NL}$ is sharper than that of $R_\text{L}$, implying nonlinear relationship between the $R_\text{L}$ and $R_\text{NL}$ as discussed later.

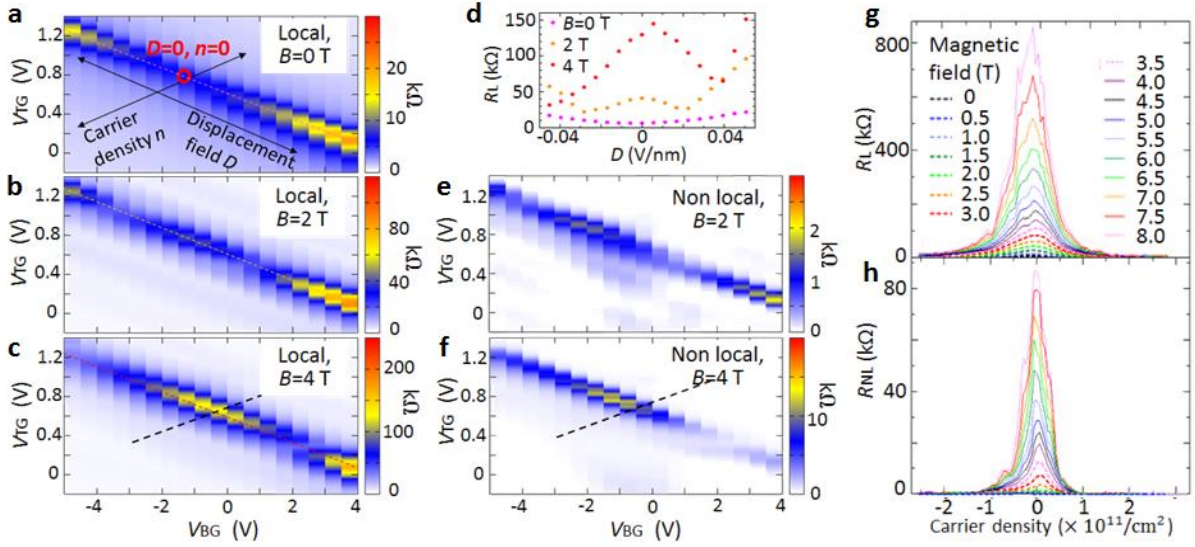

**Fig.2 | Gate voltage dependence of the local resistance ($R_\text{L}$) and nonlocal resistance ($R_\text{NL}$) in normal edge sample at $T = 1.7$ K.**

**a, b, c,** Gate voltage dependence of $R_\text{L}$ for the magnetic field $B$ of 0 T (a), 2 T (b), and 4 T (c). The red circle in the graph of $B = 0$ T shows the $D = 0$, $n = 0$ point. Dashed pink (a), orange (b), and red (c) lines indicate $n = 0$, whose cross section is shown in Fig. 2d. Dashed black line in c indicates $D = 0$, whose cross section is shown in Fig. 2g. **d,** Displacement field dependence of the $R_\text{L}$ at $n = 0$ (cross section along the dashed line of a, b, c) for the magnetic field $B$ of 0, 2, 4 T. **e, f,** Gate voltage dependence of $R_\text{NL}$ for $B = 2$ T (e), and 4 T (f). The dashed line in Fig. 2f indicates $D = 0$, whose cross section is shown in Fig. 2h. **g, h,** Carrier density dependence of the $R_\text{L}$ (g) and $R_\text{NL}$ (h) at $D = 0$ (cross section along the dashed line of c and f) for the magnetic field of 1 ~ 8 T.

Now we argue other possible origins of the nonlocal transport. First, the sign of $R_\text{NL}$ does not change between the positive and negative magnetic field (Supplementary Information Fig.1), so we can eliminate the possibility of conventional Hall voltage. Second, Ettingshausen and Nernst effect can cause nonlocal voltage [29,30] but it is difficult to explain the $R_\text{L}$ dependence of $R_\text{NL}$ discussed later (see Supplementary Information). Third, current leakage through two measurement terminals into the ground also can cause nonlocal voltage, but this is not the origin of $R_\text{NL}$ neither because input impedance of the voltage amplifiers is high enough (Supplementary Information Fig.S1b, Table S1).

Another possible origin of the nonlocal transport could be the edge transport (Fig.1f). Although it has been established both experimentally [8-15] and theoretically [2,20] that the LAF state does not have an edge channel, gate inhomogeneity at the sample edge can cause carrier doping and produces less-resistive regions along the edge. In order to eliminate the edge contribution, we employed a long-edge (Fig.1d) sample, which has a same dimension with normal edge sample except for a 10 times longer edge than the neighboring standard Hall bar. If we assume the simplest situation where the transport is only caused by the edge transport and neglect the bulk transport, $R_{\mathrm{NL}}$ should become 10 times smaller for the long-edge sample. Figure 3 shows obtained peak value of $R_{\mathrm{NL}}$ plotted as a function of $R_{\mathrm{L}}$ at the CNP under the magnetic field varied from 1 T to 8 T at various temperatures from 1.7 K to 30 K. Comparable values obtained for the two samples of different edge lengths contradict the edge contribution and are consistent with the spin-valley current in the bulk as the main contribution. In addition, we measured $R_{\mathrm{NL}}$ for other two samples with different edge lengths and obtained $R_{\mathrm{NL}}$ of the same order (see Supplementary Information).

We now turn to the relationship between the $R_{\mathrm{L}}$ and $R_{\mathrm{NL}}$ obtained by changing the magnetic field and temperature (see Figure 3). We find that all data points lie on a single curve of the cubic scaling.
Fig. 3 insets show bias voltage dependence of the differential conductance measured between terminals 1 and 3 of Fig.1e, and the gap size evaluated from it. Note energy scale of the measured temperature (1.7~ 32 K) is smaller than the gap size.

In analogy with the spin Hall effect and valley Hall effect [26,27], $R_{\mathrm{NL}}$ arising from the spin-valley Hall effect and inverse spin-valley Hall effect is given by

$$R_{\mathrm{NL}} = \frac{W}{2l} \frac{\sigma_H^2}{\sigma_{xx}(\sigma_{xx}^2 + \sigma_H^2)} \; e^{-L/l} \quad (1)$$

$$\propto \rho^3 \; (\sigma_{xx} > \sigma_H)$$
$$\propto \rho \; (\sigma_{xx} < \sigma_H)$$

for a homogeneous semiclassical system, where $\sigma_{xx}$ is local conductivity, $\sigma_H$ is spin-valley Hall conductivity, $W$ is sample width, $L$ is sample length, and $l$ is scattering length of the spin-valley current. Supposing $l$ is constant and $\sigma_H = 4e^2/h$, $R_{\mathrm{L}}$ and $R_{\mathrm{NL}}$ should have the cubic scaling relationship for $\sigma_{xx} > \sigma_H$, supporting the scenario of the spin-valley Hall effect in our experiments. The scaling changes from cubic to linear in Equation 1 for higher resistance, i.e. $\sigma_{xx} < \sigma_H$. In the measurement of

Fig.3, however, we observed cubic scaling in the entire measurement region for the long-edge sample. In the other sample of the normal-edge length with higher resistance at the CNP, we found linear scaling above $R_L \sim 90$ kΩ ($\sigma_{xx} \sim 0.7$ $e^2/h$). The reason for the cubic scaling up to higher resistivity than $h/4e^2$ is not yet clearly understood but it suggests that the effective conductivity of Eq.1 is higher than the measured conductivity, or that the effective spin-valley Hall conductivity is smaller than $4e^2/h$. Actually the cubic scaling up to the higher resistance regime was also observed in previous experiments on the valley Hall effect in bilayer graphene [26,27]. It was shown in ref. 26 that the cubic scaling holds for thermally activated component of the transport even for high $R_L$. Swirly behavior seen in fig. 3 under low magnetic field and high temperature is related to non-monotonic temperature dependence of $R_L$ and $R_{NL}$ (see Supplementary Information), whose mechanism is not yet clearly understood.

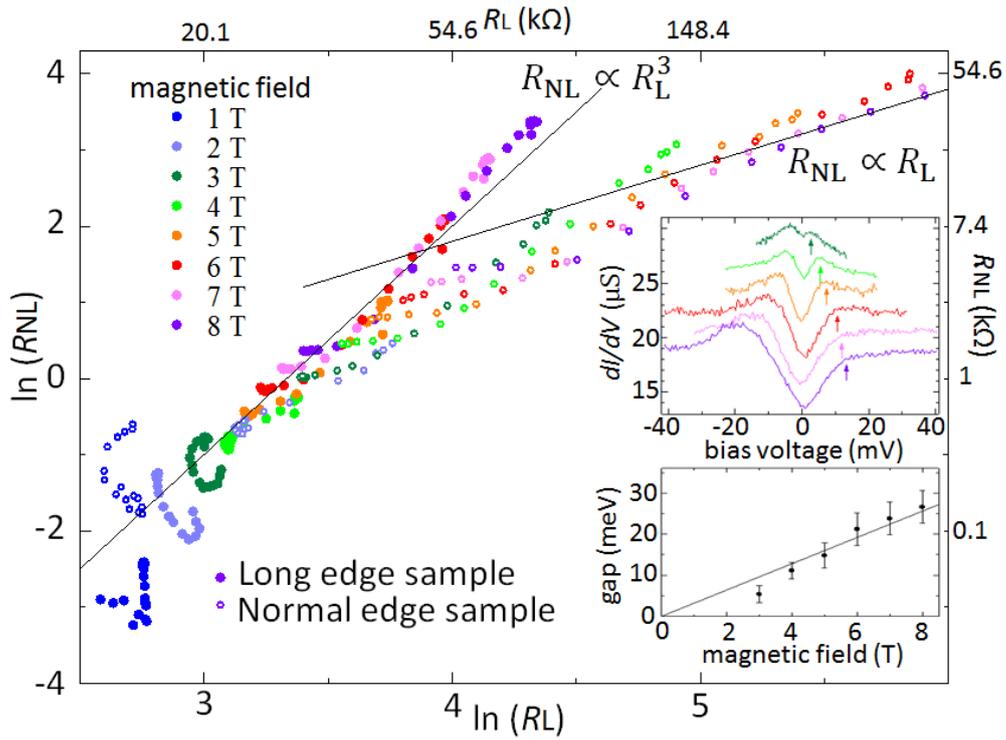

**Fig.3 | scaling relationship between $R_L$ and $R_{NL}$**

Peak value of $R_{NL}$ along $D = 0$ line plotted as a function of that of $R_L$ for temperatures 1.7 ~ 32 K. The data of the same color were taken under the same magnetic field (1 ~ 8 T). Dots and open circles denote the data from the long-edge Hall bar and the normal Hall bar, respectively. The solid lines corresponds to the cubic scaling and the linear scaling.

Inset : Upper panel is the differential conductance as a function of the bias voltage at the CNP for

$D = 0$. The lower panel shows the gap size extracted from the data of the upper panel plotted as a function of the magnetic field. All data were taken at 1.7 K.

Our demonstration evidences the spin-valley Hall effect induced by the spontaneous symmetry breaking driven by the electronic correlation. While intrinsic Hall conductivity has widely been discussed in varieties of materials, most of them are induced by symmetry breaking of crystal structures. Bilayer graphene provides a unique system, where spontaneous symmetry breaking is electrically tunable. This highly correlated state is also a spin-valley-coupled state, and largely contributing to the development of low power consumption devices utilizing non-charge degrees of freedom.

# Method

We used a mechanical exfoliation technique to prepare bilayer graphene (BLG) and h-BN flakes. The number of layers in each graphene flake on the $SiO_2$ (285 nm)/Si substrate was identified by contrast of optical microscope image. After choosing clean h-BN and graphene flakes using AFM, we stacked them. First an h-BN flake (around 30 nm thickness) was picked up by a stamp made of a polycarbonate thin film on round-shaped PDMS. Then we picked up a bilayer graphene flake by the h-BN flake, and release them on another h-BN flake. After making the h-BN/BLG/h-BN stack, we annealed it at 380 ℃ in an $Ar/H_2$ atmosphere for 1.5 hours to remove polycarbonate residue. Top gate (Pd 5 nm/Au 30 nm) and Ohmic contacts (Pd 20 nm/Au 100 nm) were defined by electron beam lithography and metal deposition by electron beam evaporators. The h-BN/BLG/h-BN stack was etched into a Hall bar by means of reactive ion etching in an $Ar/O_2/CF_4$ atmosphere.

Local and nonlocal transport experiments were performed using lock-in amplifiers (around 13 Hz) and preamplifier with input impedance of $100\,\mathrm{M\Omega}$. This input impedance is large enough for nonlocal measurement (see Supplementary Information). Bias dependence of the two-terminal differential conductance was measured using a voltage source, preamplifiers, voltmeters, and ammeters.

# Supplemental information

## I  Elimination of other possible origins of the nonlocal transport

Here we discuss and eliminate the possibility of the normal Hall voltage, Ettingshausen and Nernst effect, and current flow into measurement terminals as an origin of nonlocal transport.

[Hall voltage]

If measurement terminals are shifted from the vertical direction to the longitudinal direction of the Hall bar, Hall voltage appears. But this normal Hall voltage is not the origin of $R_{\mathrm{NL}}$ in our device because $R_{\mathrm{NL}}$ becomes maximal at zero carrier density. In addition, we confirmed that sign of $R_{\mathrm{NL}}$ is not altered by reversing the magnetic field (Fig. S1a,b). This result is inconsistent with the normal Hall voltage while it is consistent with the scenario of the spin-valley Hall effect.

[Ettingdhausen and Nernst effect]

Ettingshausen effect is the magnetic analogue of Peltier effect where thermal gradient appears perpendicular to the magnetic field and charge current. Heat flow generated by the Ettingshausen effect causes nonlocal voltage due to Nernst effect (Fig. S1c) [29,30]. The Ettingshausen and Nernst effects are described by the following formulas:

$$\bm{J} = K_0 \bm{E} + \frac{K_1}{eT} \nabla T \quad (S1)$$

$$\bm{J}_Q = -\frac{K_1}{e}\bm{E} - \frac{K_2}{e^2 T}\nabla T \quad (S2)$$

$$K_n \equiv \int (\varepsilon - \mu)^n \sigma(\varepsilon, T) \frac{\partial f}{\partial \varepsilon} d\varepsilon.$$

Here, $\bm{J}$ is the charge current, $\bm{J}_Q$ is the heat flow, $\bm{E}$ is the electric field, $T$ is the temperature, $\sigma$ is the spectrum function of conductivity matrix, and $f$ is Fermi distribution function. We assume $\bm{J}, \bm{E}$ in y direction and $\bm{J}_Q, \nabla T$ in x direction. Using the electrical conductivity $\sigma_{xx} = K_{0\,xx}$, Nernst coefficient $S_{yx} = K_{1\,xy}/eTK_{0\,xx}$, and the electron thermal conductivity $\kappa = -K_{2\,xx}/e^2 T$, Eqs. S1 and S2 are written as

$$\begin{pmatrix} J_y \\ J_{Qx} \end{pmatrix} = \begin{pmatrix} \sigma_{xx} & \sigma_{xx} S_{yx} \\ -T\sigma_{xx} S_{yx} & \kappa \end{pmatrix} \begin{pmatrix} E_y \\ \nabla_x T \end{pmatrix}. \quad (S3)$$

The heat flow $Q$ generated by the current injection and Ettingshausen effect is derived from Eq. S3 as follows,

$$J_{Qx} = -T\sigma_{xx} S_{yx} E_y = S_{yx} T J_y. \quad (S4)$$

Nonlocal electric field between the detection terminals induced by the Nernst effect is derived from inverse matrix of Eq. S3 and becomes

$$E_{\text{NL}} = \frac{-\sigma_{xx} S_{yx} J_{Qx} \gamma}{\sigma_{xx}\kappa + T(\sigma_{xx}S_{yx})^2} = \frac{-\sigma_{xx} S_{yx}^2 T J_y \gamma}{\sigma_{xx}\kappa + T(\sigma_{xx}S_{yx})^2}, \quad (S5)$$

where $\gamma$ is decay rate of the heat flow. Therefore nonlocal resistance is given as

$$R_{\text{NL}} \propto \frac{S_{yx}^2 T}{\kappa} \quad (\kappa/T\sigma_{xx} \gg S_{yx}^2) \quad (S6)$$

$$\propto \frac{1}{\sigma_{xx}} \quad (\kappa/T\sigma_{xx} \ll S_{yx}^2) \quad (S7)$$

It has been experimentally confirmed that $S_{yx}$ at the CNP of bilayer graphene is proportional to $T$ [ref. S1]. If $R_{\text{NL}}$ is due to the Ettingshausen and Nernst effect, we should assume the electron thermal conductivity as $\kappa \propto T^3 \rho^{-3}$ to account for the cubic scaling relationship between $R_{\text{NL}}$ and $R_{\text{L}}$. However this contradicts to Wiedemann–Franz law $\kappa = (\pi^2/3)(k_B/e)^2 T \sigma_{xx}$. Therefore it is unlikely that Ettingshausen and Nernst effect play the major role in the observed nonlocal transport.

In addition, $R_{\text{NL}}$ expected from the Ettingshausen and Nernst effect is much smaller than the measured value. In ref. 29, $R_{\text{NL,Ettingshausen}}$ at the CNP of monolayer graphene under the magnetic field is estimated to be $10\,\Omega$ for $\rho_{xx}\sim 10\,\text{k}\Omega$ at $T = 4.2\,\text{K}$, so $R_{\text{NL,Ettingshausen}}/\rho_{xx}T^2 \sim 10^{-5}\,(\text{K}^{-2})$. For $\rho_{xx}\sim 10\,\text{k}\Omega$, $R_{\text{NL}}$ in our measurement is about $1\,\text{k}\Omega$ at $T\sim 20\,\text{K}$, so $R_{\text{NL}}/\rho_{xx}T^2 \sim 10^{-4}\,(\text{K}^{-2})$. Sample dimension is almost the same between ref. 29 and ours. Although we use bilayer graphene encapsulated by h-BN while they use graphene on SiO$_2$ substrate, h-BN may only reduce $R_{\text{NL}}$, because larger thermal conductivity of h-BN than that of SiO$_2$ enhances the dissipation of heat current flowing along the Hall bar.

[Current flow into measurement terminals]
Our measurement was performed in the highly resistive regime of the CNP. When the resistance of the sample is high at the CNP under a high magnetic field, an electrical current flows into the voltage amplifiers that are connected to the two measurement terminals (Fig. S1d). $R_{\text{NL}}$ originated from this effect is given by

$$R_{\text{NL,amp}} = \frac{rR}{R + R_{\text{amp}}/2} \lesssim \frac{2rR}{R_{\text{amp}}} \quad (S10)$$

Where $R$ is the resistance between the current injection terminals, $r$ is difference of the contact resistance between the two measurement terminals obtained from the two-terminal measurement, and $R_{amp}$=100 MOhm is the input impedance of the voltage amplifiers.

Table S1 shows measured $R_{NL}$ and calculated $R_{NL,amp}$ by eq.S10 for several data points. $R_{NL,amp}$ is much smaller than obtained $R_{NL}$. The current flow into the voltage amplifiers is therefore not the origin of the observed nonlocal resistance.

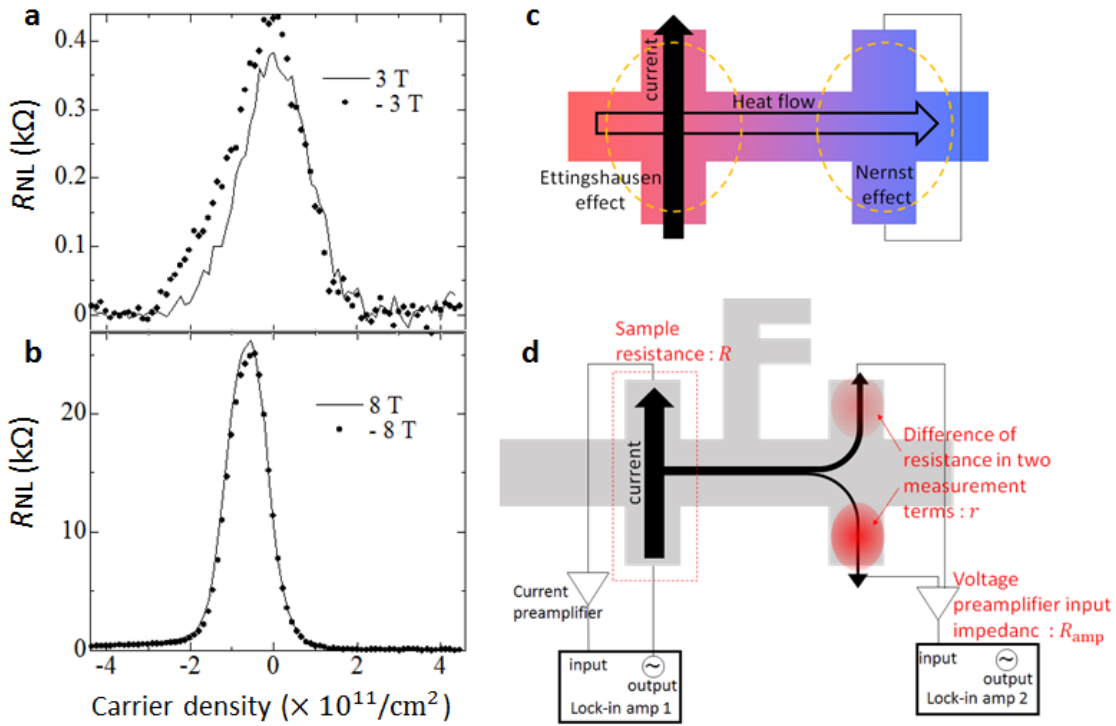

**Fig.S1 | $R_{NL}$ under reversed magnetic field and current flow into measurement terminals.**
**a,b,** Carrier density dependence of $R_{NL}$ at $D = 0$ for reversed magnetic fields of 3 (-3) T (a) and 8 (-8) T (b). The data is taken at 1.67 K. **c,** Schematic description of nonlocal transport by Ettingshausen and Nernst effect. **d,** Schematic description of the current flow into the voltage amplifiers

| Measured $R_L$ | Measured $R_{NL}$ | R | r | $R_{amp}$ | $R_{NL, amp}$ |
|---|---|---|---|---|---|
| 20 kΩ | 367 Ω | 20 kΩ | < 20 kΩ | 100 MΩ | < 8 Ω |
| 70 kΩ | 20 kΩ | 70 kΩ | < 70 kΩ | 100 MΩ | < 98 Ω |
| 300 kΩ | 33 kΩ | 300 kΩ | < 300 kΩ | 100 MΩ | < 1.8 kΩ |

**Table.S1 | Estimated resistance due to current flow into measurement terminals.**
$R_L$ and $R_{NL}$ extracted from Fig.3 of main text, estimated values of $R(=R_L)$ and $r\,(<R)$, $R_{amp}$, and calculated values of $R_{NL,amp}$ using eq. S9.

## II Sample dependence

In the main text, we compared $R_{NL}$ of two samples with different edge lengths. Here, we show the data of other two samples (Fig. S2a, sample3, 4). They have the same order of $R_{NL}$ is spite of large difference of the edge length, which is consistent with the scenario of the bulk spin-valley current.

However their peak positions of $R_{NL}$ as functions of the gate voltages were slightly shifted from those of $R_L$ (Fig. S2b) and therefore unsuitable for further analysis. This shift most probably originates from inhomogeneity of the carrier density.

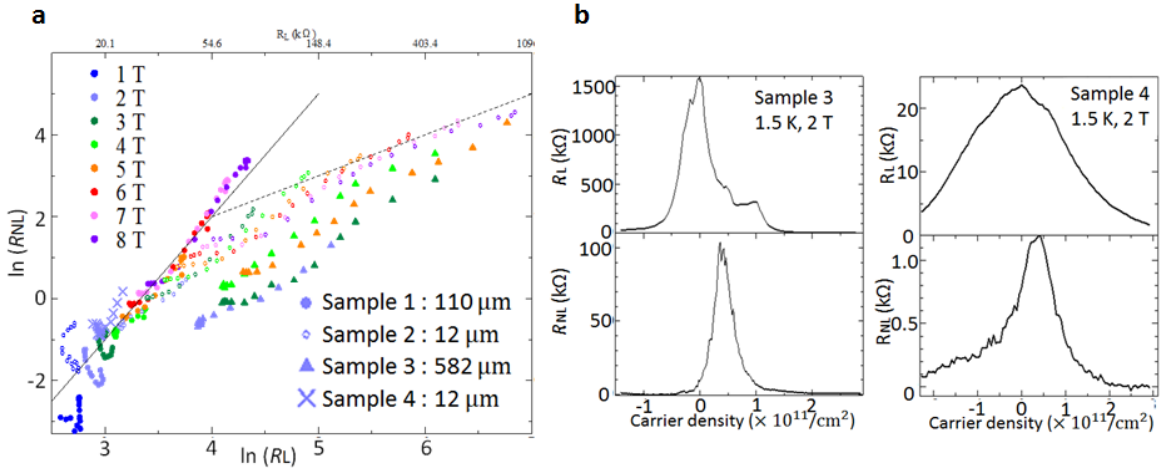

**Fig. S2 | Sample dependence of $R_L$ and $R_{NL}$.**
**a,** Peak value of the $R_{NL}$ along D = 0 line plotted as a function of that of the $R_{NL}$ for the temperatures 1.7 ~ 32 K. The data of the same color were taken under the same magnetic field (1 ~ 8 T). Dots, open circles, triangles, and crosses denote the data from different samples. Edge lengths of the samples 1~4 are 110 μm, 12 μm, 582 μm, and 12 μm, respectively. Sample1 is the long-edge sample and sample2 is the normal-edge sample in the main text. **b,** Carrier density dependence of the $R_L$ and $R_{NL}$ at $D = 0$ at 1.5 K for the magnetic field of 2 T in sample 3 and 4.

## III Temperature dependence in high temperature region

We show temperature dependence of $R_\text{L}$ and $R_\text{NL}$ for $T = 1.7 \sim 280$ K (Fig. S3a-d). $R_\text{L}$ and $R_\text{NL}$ do not depend monotonically on the temperature. For example, by increasing the temperature, the sample changes to be metallic around $T = 8$ K and again becomes insulating around $T = 26$ K for $B = 2$ T (see slope of Fig. S3a). We define $T_{c1(2)}$ as the temperature at which slope of the resistance as a function of $T$ changes from negative (positive) to positive (negative), and plot them in Fig. S3f.

$T_{c1}$ of $R_\text{L}$ ($T_{c1,\text{local}}$) and that of $R_\text{NL}$ ($T_{c1,\text{nonlocal}}$) increase with the magnetic field. $T_{c1,\text{local}}$ is smaller than $T_{c1,\text{nonlocal}}$ for $B \leq 4$ T, but larger for $B \geq 6$ T.

$T_{c2}$ of $R_\text{L}$ ($T_{c2,\text{local}}$) is almost linear with $B$ and the proportionality factor is similar to that of the gap size estimated from bias measurement (main text Fig.3 inset), although we cannot define $T_{c2,\text{local}}$ for $B \geq 5$ T. $T_{c2}$ of $R_\text{NL}$ ($T_{c2,\text{nonlocal}}$) is larger than $T_{c2,\text{local}}$.

In addition, we found that cubic scaling relationship between $R_\text{L}$ and $R_\text{NL}$ does not hold for $T > T_{c1,\text{local}}$.

We suppose that decrease of $R_\text{L}$ above $T_{c2,\text{local}}$ is related to the breaking of the spin order of LAF state, whose robustness is characterized by the gap size, because $T_{c2,\text{local}}$ is comparable with the gap size. However, some features are not yet understood such as the metallic temperature dependence for $T_{c1} < T < T_{c2}$, and difference between $T_{c,\text{local}}$ and $T_{c,\text{nonlocal}}$. Further theoretical and experimental investigation is necessary to account for the temperature dependence of $R_\text{L}$ and $R_\text{NL}$. In particular, experiments using spin injection or NMR may be useful to reveal the breaking of the spin order of LAF state at high temperature.

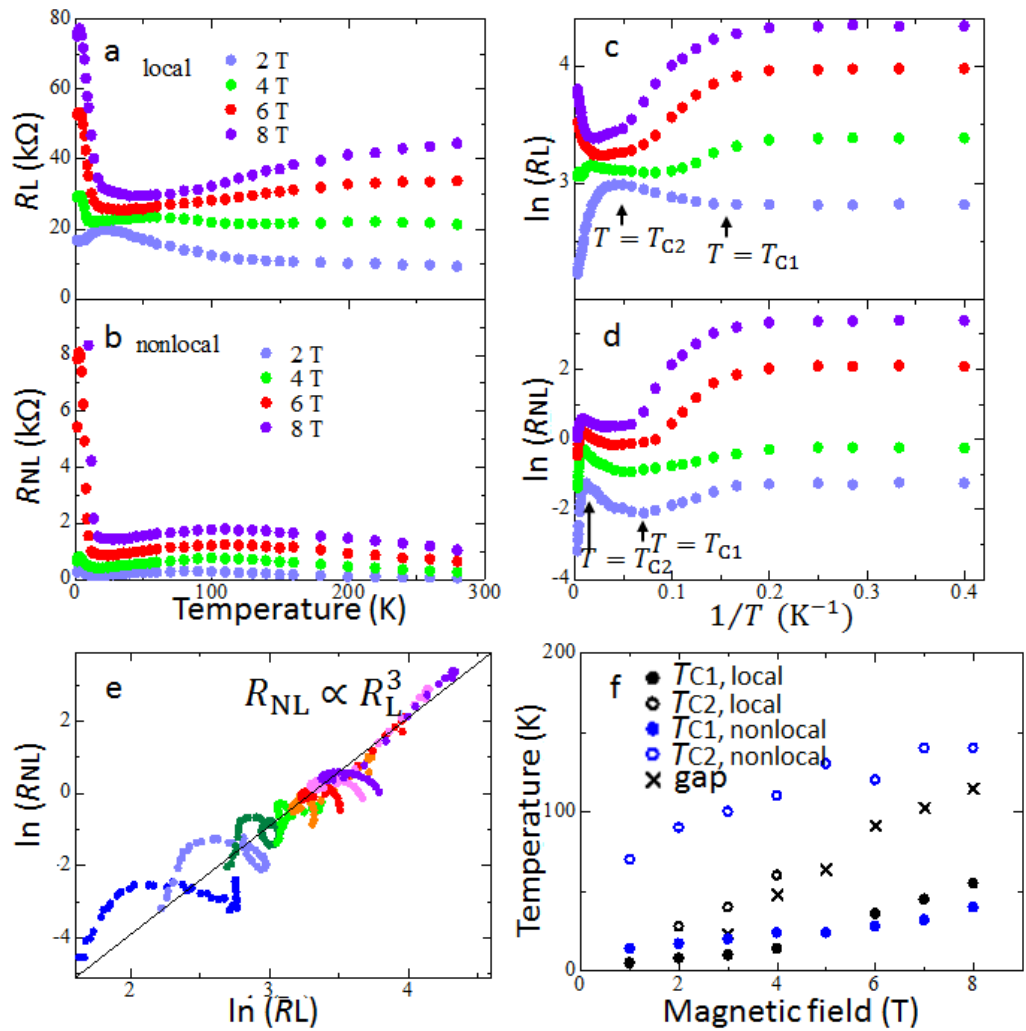

**Fig.S3 | Temperature dependence of $R_L$ and $R_{NL}$ in high temperature region**

**a, b** Temperature dependence of $R_L$ (a) and $R_{NL}$ (b) at $D = 0$, $n = 0$ in sample 1 (long edge sample in main text). **c, d** Alenius plot of a (c) and b (d). Arrows indicate examples of $T_{c1}$ and $T_{c2}$. **e,** Relationship between $R_L$ and $R_{NL}$ of sample 1 for the temperature between 1.7K and 280 K. The solid line corresponds to the cubic scaling. **f,** Magnetic field dependence of $T_c$ (indicated as arrows in Fig. S3b) and the gap size estimated from bias measurement.

Reference

[S1] Wang, C. R. et al. Transverse thermoelectric conductivity of bilayer graphene in the quantum Hall regime. *Phys. Rev. B* 82, 121406 (2010)